\documentclass[12pt,floatfix]{revtex4}
\usepackage{graphicx}
\usepackage{amsmath}
\usepackage{amssymb}
\usepackage[usenames,dvipsnames]{color}

\begin{document}

\title{$\alpha$-helix$\leftrightarrow$random
coil phase transition: analysis of {\it ab initio} theory
predictions}

\author{Ilia A. Solov'yov$^{\text *}$, Alexander V. Yakubovich$^{\text *}$, Andrey V. Solov'yov\footnote{On leave from the A.F. Ioffe
Institute, St. Petersburg, Russia. E-mail:
ilia@fias.uni-frankfurt.de} and Walter Greiner}

\affiliation{Frankfurt Institute for Advanced Studies, Max von Laue
Str. 1, 60438 Frankfurt am Main, Germany}

\begin{abstract}


In the present paper we present results of calculations obtained
with the use of the theoretical method described in our preceding
paper \cite{Yakubovich07_preceding} and perform detail analysis of
$\alpha$-helix$\leftrightarrow$random coil transition in alanine
polypeptides of different length. We have calculated the potential
energy surfaces of polypeptides with respect to their twisting
degrees of freedom and construct a parameter-free partition function
of the polypeptide using the suggested method
\cite{Yakubovich07_preceding}. From the build up partition function
we derive various thermodynamical characteristics for alanine
polypeptides of different length as a function of temperature. Thus,
we analyze the temperature dependence of the heat capacity, latent
heat and helicity for alanine polypeptides consisting of 21, 30, 40,
50 and 100 amino acids. Alternatively, we have obtained same
thermodynamical characteristics from the use of molecular dynamics
simulations and compared them with the results of the new
statistical mechanics approach. The comparison proves the validity
of the statistical mechanic approach and establishes its accuracy.
\end{abstract}


\maketitle

\section{Introduction}
\label{intro}

In our preceding paper \cite{Yakubovich07_preceding}, we introduced
a novel and general theoretical method for the description of phase
transitions in finite complex molecular systems. In particular, we
have demonstrated that for polypeptide chains, i.e. chains of amino
acids, one can identify specific twisting degrees of freedom that
are responsible for the folding dynamics of these amino acid chains.
In other words, these degrees of freedom characterize the transition
from a chain in a random coil state, to one in an $\alpha$-helix
structure and vice versa.

The essential domains of the potential energy surface (PES) of
polypeptides with respect to these twisting degrees of freedom have
been calculated and thorougly analyzed on the basis of both
classical molecular dynamics (MD) simulations, and ab initio methods
such as density functional theory (DFT) and the Hartree-Fock
approach. In
Refs.~\cite{Yakubovich07_preceding,Yakubovich06a,Yakubovich06a_EPN},
it was shown that with the PES, one can
construct a partition function of a polypeptide chain from which it
is then possible to extract all essential thermodynamical variables
and properties, such as the heat capacity, phase transition
temperature, free energy, etc.

In this paper, we explore this further using a formalism we
introduced previously \cite{Yakubovich07_preceding} and apply it to
a detailed analysis of the $\alpha$-helix$\leftrightarrow$random
coil phase transition in alanine polypeptides of different lengths.
We have chosen this system because it has been widely investigated
both theoretically
\cite{Zimm59,Gibbs59,Lifson61,Schellman58,Lifson64,Poland66a,Ooi91,Gomez95,Tobias91,Garcia02,Nymeyer03,Irbaeck04,Shental-Bechor05,Kromhout01,Chakrabartty94,Scheraga70,Scheraga02}
and experimentally \cite{Scholtz91,Lednev01,Thompson97,Williams96}
during the last five decades (for review see, e.g.
\cite{Shakhnovich06,Ptizin_book,Shea01,Prabhu05}) and thus is
perfect system for testing a novel theoretical approach.

The theoretical studies of the helix-coil transition in polypeptides
have been performed both with the use of statistical mechanics
methods
\cite{Zimm59,Gibbs59,Lifson61,Schellman58,Lifson64,Poland66a,Kromhout01,Chakrabartty94,Shea01,Scheraga70,Scheraga02,Shental-Bechor05}
and of MD
\cite{Tobias91,Garcia02,Nymeyer03,Irbaeck04,Shental-Bechor05}.
Previous attempts to describe the helix-coil transition in
polypeptide chains using the principles of of statistical mechanics
were based on the models suggested in sixties
\cite{Zimm59,Gibbs59,Lifson61,Schellman58}. These models were based
on the construction of the polypeptide partition function depending
on several parameters and were widely used in
Refs.~\cite{Kromhout01,Chakrabartty94,Shakhnovich06,Ptizin_book,Shea01,Scheraga70,Scheraga02,Shental-Bechor05}
for the description of the helix-coil transition in polypeptides.

For a comprehensive overview of the relevant work we refer to recent
reviews \cite{Shakhnovich06,Shea01,Prabhu05} and the book
\cite{Ptizin_book}, as well as to our preceding paper
\cite{Yakubovich07_preceding}.

Experimentally, extensive studies of the helix-coil transition in
polypeptides have been conducted
\cite{Scholtz91,Lednev01,Thompson97,Williams96}. In
Ref.~\cite{Scholtz91}, the enthalpy change of an $\alpha$-helix to
random coil transition for the Ac-Y(AEAAKA)$_8$F-NH$_2$ peptide in
water was determined calorimetrically. The dependence of the heat
capacity of the polypeptide on temperature was measured using
differential scanning calorimetry. In
Refs.~\cite{Lednev01,Thompson97}, UV resonance Raman spectroscopy
was performed on the MABA-[A]$_5$-[AAARA]$_3$-ANH$_2$ peptide. Using
circular dichroism methods, the dependence of helicity on
temperature was measured. While in Ref.~\cite{Williams96}, the
kinetics of the helix-coil transition of the 21-residue alanine
polypeptide was investigated by means of infrared spectroscopy.

In this work, we have calculated the PES
of polyalanines of different lengths with respect to their twisting
degrees of freedom. This was done within the framework of classical
molecular mechanics. However, to scrutinize the accuracy of these
calculations, we compared the resultant molecular mechanics potential energy
landscapes with those obtained using {\it ab initio} density
functional theory (DFT). The comparison was only performed for
alanine tripeptide and hexapeptide, since for larger polypeptides,
the DFT calculation becomes increasingly computationally demanding.
Hence for these larger systems, only molecular mechanics simulations
have been used in this present work.

The calculated PES was then used to construct a
parameter-free partition function of the polypeptide using the
statistical method we had outlined in our preceding paper
\cite{Yakubovich07_preceding}. This partition function was then used
to derive various thermodynamical characteristics of alanine
polypeptides as a function of temperature and polypeptide length. We
have calculated and analyzed the temperature dependence of the heat
capacity, latent heat and helicity of alanine polypeptides
consisting of 21, 30, 40, 50 and 100 amino acids. We have also
established a correspondence between our {\it ab initio} method with
the results of the semiempirical approach of Zimm and Bragg
\cite{Zimm59}. Thus, on the basis of our approach, we have
determined the key parameters of the Zimm-Bragg theory that itself
utilizes principles of statistical mechanics.

Finally, we have calculated the heat capacity, latent heat and
helicity of alanine polypeptides using molecular dynamics and have
compared the obtained results with those using our statistical
approach. Comparison between the two methods allows us to establish
the accuracy of our statistical method for relatively small
molecular systems, and lets us gauge the feasibility of extending
the description to larger molecular objects for which it is
especially essential in those cases where MD simulations are hardly
possible due to computational limitations.

Our paper is organized as follows.  In section \ref{theory} we
present the final expressions obtained within the formalism
described in our preceding paper \cite{Yakubovich07_preceding} and
introduce basic equations and the set of parameters which have been
used in MD calculations. In section \ref{results} we present and
discuss the results of computer simulations obtained with the use of
developed theoretical method and compare then with results of MD
simulations. In section \ref{conclusion}, we draw a conclusion to
this paper.

\section{Theoretical methods}
\label{theory}

\subsection{Statistical model for the
$\alpha$-helix$\leftrightarrow$random coil phase transition}

Our calculations have been performed using the statistical formalism we described previously
\cite{Yakubovich07_preceding}. Here, we will only outline the
basic ideas of this method and present the final expressions that were used in our investigation.

Let us consider a polypeptide, consisting of $n$ amino acids. The
polypeptide can be found in one of its numerous isomeric states with
different energies. A group of isomeric states with similar
characteristic physical properties is called a {\it phase state} of
the polypeptide. Thus, a regular bounded $\alpha$-helix state
corresponds to one phase state of the polypeptide, while all
possible unbounded random conformations can be denoted as the random
coil phase state.

The {\it phase transition} is then a transformation of the polypeptide
from one phase state to another, i.e. the transition from a regular
$\alpha$-helix to a random coil conformation.

All thermodynamical properties of a molecular system are
described by its partition function. The partition function of a
polypeptide can be expressed as
\cite{Yakubovich07_preceding}):

\begin{eqnarray}
\nonumber {\mathbb Z}&=&A\cdot
B(kT)\cdot(kT)^{3N-3-\frac{l_s}{2}}\left[\beta
Z_{b}^{n-1}Z_{u}+\beta\sum_{i=1}^{n-4}(i+1)Z_{b}^{n-i-1}Z_{u}^{i+1}+Z_{u}^n+\right.\\
&&\left.+\sum_{i=2}^{(n-3)/2}\beta^i\sum_{k=i}^{n-i-3}\frac{(k-1)!(n-k-3)!}{i!(i-1)!(k-i)!(n-k-i-3)!}Z_b^{k+3i}Z_u^{n-k-3i}\right]
\label{Zhc}
\end{eqnarray}

\noindent Here the first and the third terms in the square brackets
describe, respectively, the partition function of the polypeptide in
the $\alpha$-helix and the random coil phases. The second term in
the square brackets accounts for the situation of phase
co-existence. The summation in this term is performed up to $n-4$ as
the shortest $\alpha$-helix has only 4 amino acids. The final term
in the square brackets accounts for the polypeptide conformations in
which a number of amino acids in the $\alpha$-helix conformation are
separated by amino acids in the random coil conformation. The first
summation in this term goes over the separated helical fragments of
the polypeptide, while the second summation goes over individual
amino acids in the corresponding fragment. Polypeptide conformations
with two or more helical fragments are energetically unfavorable.
This fact will be discussed in detail further on in this paper.
Therefore, the fourth term in the square brackets Eq.~(\ref{Zhc})
can be omitted in the construction of the partition function.

\begin{figure}[h]
\includegraphics[scale=0.8,clip]{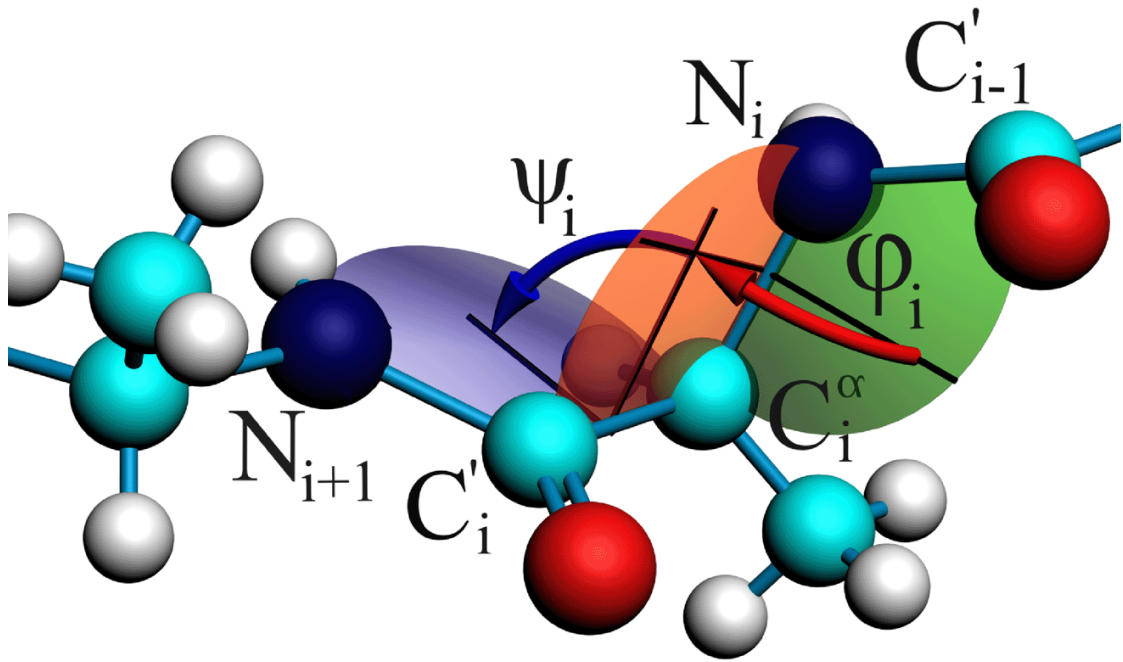}
\caption{Dihedral angles $\varphi$ and $\psi$ used for
characterization of the secondary structure of a polypeptide chain.}
\label{fg:angle_def}
\end{figure}

$A$ in Eq.~(\ref{Zhc}) is a factor that is determined by the
specific volume, momenta of inertia and frequencies of normal
vibration modes of the polypeptide in different conformations
\cite{Yakubovich07_preceding}, $l_s$ is the total number of the
"soft" degrees of freedom in the system. $B(kT)$ is a function
defined in our preceding paper \cite{Yakubovich07_preceding}, which
describes the rotation of the side radicals in the polypeptide.
$Z_b$ and $Z_u$ are the contributions to the partition function from
a single amino acid being in the bounded or unbounded states
respectively. They can be written as:

\begin{eqnarray}
\label{Zb} Z_b&=&\int_{-\pi}^{\pi }\int_{-\pi}^{ \pi}
\exp\left({-\frac{\epsilon^{(b)}(\varphi,\psi)}{kT}}\right){\rm
d}\varphi{\rm d}\psi\\
\label{Zu} Z_u&=&\int_{-\pi}^{\pi }\int_{-\pi}^{ \pi}
\exp\left({-\frac{\epsilon^{(u)}(\varphi,\psi)}{kT}}\right){\rm
d}\varphi{\rm d}\psi\\
\label{beta} \beta&=&\left(\int_{-\pi}^{\pi }\int_{-\pi}^{ \pi}
\exp\left({-\frac{\epsilon^{(b)}(\varphi,\psi)+\epsilon^{(u)}(\varphi,\psi)}{kT}}\right){\rm
d}\varphi{\rm d}\psi\right)^3,
\end{eqnarray}

\noindent where $k$ and $T$ are the Boltzmann constant and the
temperature respectively, while $N$ is the total number of atoms in
the system. $\epsilon^{(b)}(\varphi,\psi)$ and
$\epsilon^{(u)}(\varphi,\psi)$ in Eqs.~(\ref{Zb})-(\ref{beta}) are
the potential energies of a single amino acid in the bounded and
unbounded conformations calculated respectively versus the twisting
degrees of freedom $\varphi$ and $\psi$. These degrees of freedom
are defined for each amino acid of the polypeptide except for the
boundary ones and are described by two dihedral angels $\varphi_i$
and $\psi_i$ (see Fig. \ref{fg:angle_def})

Both angles are defined by four neighboring atoms in the polypeptide
chain. The angle $\varphi_i$ is defined as the dihedral angle
between the planes formed by the atoms
($C_{i-1}^{'}-N_{i}-C_i^{\alpha}$) and
($N_{i}-C_i^{\alpha}-C_i^{'}$). While the angle $\psi_i$ is defined
as the dihedral angle between the ($N_{i}-C_{i}^{\alpha}-C_i^{'}$)
and ($C_{i}^{\alpha}-C_i^{'}-N_{i+1}$) planes. The atoms are
numbered from the NH$_2$- terminal of the polypeptide and
$\varphi_i$ and $\psi_i$ take all possible values within the
interval [$-180^{\circ}$;$180^{\circ}$]. For an unambiguous
definition most commonly
used\cite{Rubin04,Yakubovich06b,Yakubovich06c,ISolovyov06b,ISolovyov06c},
$\varphi_i$ and $\psi_i$ are counted clockwise if one looks on the
molecule from its NH$_2$- terminal (see Fig.\ref{fg:angle_def}).

By substituting Eqs.~(\ref{Zb}) and (\ref{Zu}) into Eq.~(\ref{Zhc}), one
obtains the final expression for the partition function of a
polypeptide experiencing an $\alpha$-helix$\leftrightarrow$random
coil phase transition. This is the expression which we then use to
evaluate all thermodynamical characteristics of our polypeptide system.

\subsection{Molecular dynamics}

Molecular dynamics (MD) is an alternative approach which can be used
for the study of phase transitions in macromolecular systems. Within
the framework of MD, one tries to solve the equations of motion for
all particles in the system interacting via a given potential. Since
the technique of MD is well known and described in numerous
textbooks~\cite{Rapaport_Book,NAMD,MD_Frenkel}, we will only present
the basic equations and ideas underlying this method.

MD simulations usually imply the numerical solution of the
Langevin equation \cite{LangEq_book,Reif_book,MD_Frenkel}:

\begin{equation}
m_i{\bf a_i}=m_i{\ddot {\bf r_i}}=-\frac{\partial U({\bf
R})}{\partial {\bf r_i}}-\beta_i{\bf v_i}+{\bf \eta}(t).
\label{Langevin}
\end{equation}

\noindent Here $m_i$, ${\bf r_i}$, ${\bf v_i}$ and ${\bf a_i}$ are
the mass, radius vector, velocity and acceleration of the atom $i$.
$U({\bf R})$ is the potential energy of the system. The second term
describes the viscous force which is proportional to the particle
velocity. The proportionality constant $\beta_i=m_i\gamma$, where
$\gamma$ is the damping coefficient. The third term is the noise
term that represents the effect of a continuous series of collisions
of the molecule with the atoms in the medium. To study the
time-evaluation of the system, the Langevin equations of motion,
Eq.~(\ref{Langevin}), are integrated for each particle.

In this paper, we use the MD approach to study the
$\alpha$-helix$\leftrightarrow$random coil phase transition in
alanine polypeptides and compare the results with those obtained using the
statistical mechanics approach. For the simulations, we use the
CHARMM27 force field \cite{CHARMM} to describe the interactions
between atoms. This is a common empirical field for treating polypeptides,
proteins and lipids\cite{CHARMM,ElsaPoster,ElsaPaper,Sotomayor05,Gullingsrud04}.

MD simulations allow one to study the
$\alpha$-helix$\leftrightarrow$random coil phase transition of
alanine polypeptide as this process occurs on the nanosecond time
scale. From these simulations, one can obtain the important characteristics of the phase transition, such
as the transition temperature, maximal heat capacity, the
temperature range of the transition and the latent heat.

We perform MD simulations of alanine polypeptides consisting of 21,
30, 40, 50 and 100 amino acids. For this study it is necessary to
specify the initial conditions for the system, i.e. to define the
initial positions of all atoms and set their initial velocities. We
assume the initial structure of the polypeptides as an ideal
$\alpha$-helix
\cite{Lehninger_Biochemistry,Ptizin_book,Biochemistry_book} and
assign the particle velocities randomly according to the Maxwell
distribution at a given temperature.

The MD simulations of the polypeptides were performed at different
temperatures. For an alanine polypeptide consisting of 21 amino
acids, 71 simulations were performed for the temperatures in the
region of 300 K$^\circ$ to 1000 K$^\circ$. For polypeptides
consisting of 30, 40, 50 and 100 amino acids, 31 simulations were
performed for each polypeptide in the temperature region of 300
K$^\circ$ to 900 K$^\circ$. The simulations were carried out within
a time interval of 100 ns and an integration step of 2 fs. The first
25 ns of the simulation were used to equilibrate the system, while
the next 75 ns were used for obtaining data about the energy and
structure of the system at a given temperature.

The set of the parameters used in our simulations can be found in
Refs. \cite{Rapaport_Book,NAMD,MD_Frenkel}. All simulations were
performed using the NAMD molecular dynamics program\cite{NAMD},
while visualization of the results was done with VMD\cite{VMD}. The
covalent bonds involving hydrogen atoms were considered as rigid.
The damping coefficient $\gamma$ was set to 5 ps$^{-1}$. The
simulations were performed in the $NVT$ canonical ensemble using a
Langevin thermostat with no cutoff for electrostatic interactions.

\section{Results  and Discussion}
\label{results}

In this section we present the results of calculations obtained
using our statistical mechanics approach and those from our MD
simulations. In subsection \ref{accuracy} we discuss the accuracy of
this force field as applied to alanine polypeptides. In subsection
\ref{potenergy} we present the PESs for
different amino acids in alanine polypeptide calculated versus the
twisting degrees of freedom $\varphi$ and $\psi$ (see
Fig.~\ref{fg:angle_def}). In subsection \ref{phase_trans}, the
statistical mechanics approach is used for the description of the
$\alpha$- helix$\leftrightarrow$random coil phase transition. Here,
the results of the statistical mechanics approach are compared to
those obtained from MD simulations. In subsection
\ref{sec:correlation} the statistical independence of amino acids in
the polypeptide is discussed.

\subsection{Accuracy of the molecular mechanics potential}
\label{accuracy}

The PES of alanine polypeptides was
calculated using the CHARMM27 force field \cite{CHARMM} that has
been parameterized for the description of proteins, in particular
polypeptides, and lipids. Nevertheless, the level of its accuracy
when applied to alanine polypeptides cannot be taken for granted and
has to be investigated. Therefore, we compare the PESs for alanine tri- and hexapeptide calculated using the
CHARMM27 force field with those calculated using {\it ab initio}
density functional theory (DFT). In the DFT approach, the PES of
alanine tri- and hexapeptides were calculated as a function of the
twisting degrees of freedom, $\varphi$ and $\psi$ (see
Fig.~\ref{fg:angle_def}), in the central amino acid of the
polypeptide~\cite{ISolovyov06b}. All other degrees of freedom were
frozen.

\begin{figure}[h]
\includegraphics[scale=0.8,clip]{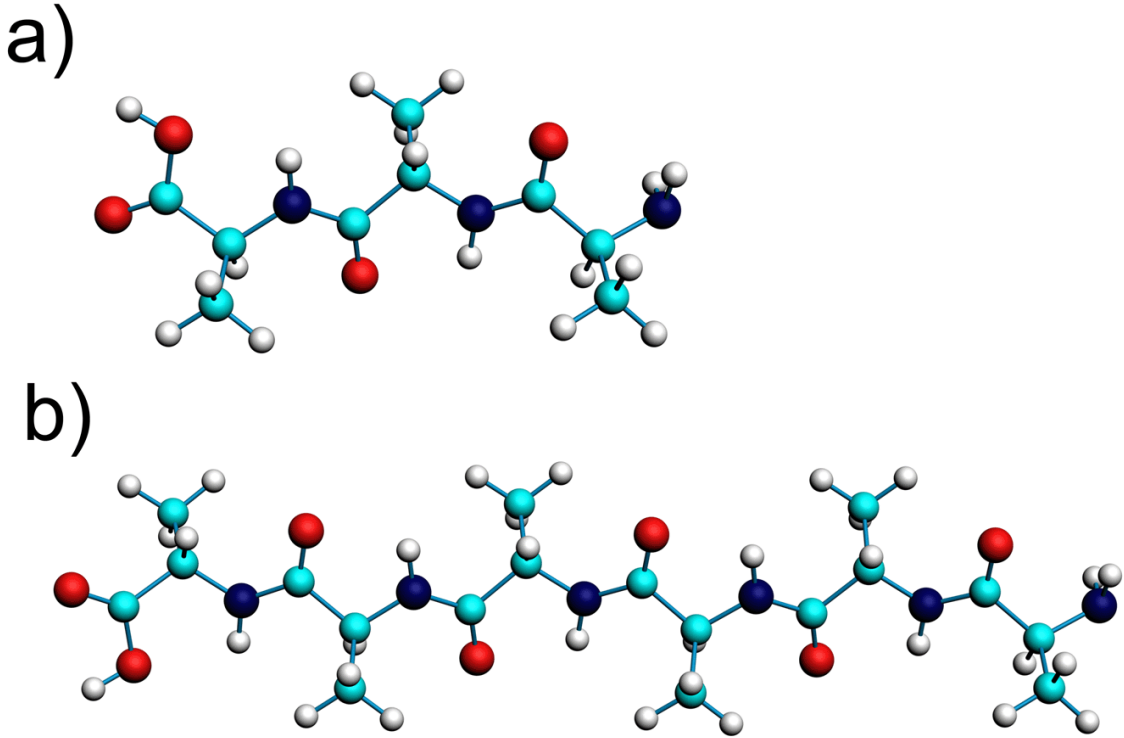}
\caption{Optimized geometries of alanine polypeptide chains: a)
Alanine tripeptide; b) Alanine hexapeptide in the $\beta$-sheet
conformation.} \label{fg:stable_geom}
\end{figure}

To establish the accuracy of the CHARMM27 force field, we have
calculated the PESs of alanine polypeptides in
its $\beta$-sheet conformation. The geometry of alanine tri- and
hexapeptide used in the calculations are shown in
Fig.~\ref{fg:stable_geom}a and Fig.~\ref{fg:stable_geom}b
respectively. The {\it ab initio} calculations were
performed\cite{ISolovyov06b} using B3LYP, Becke's three-parameter
gradient-corrected exchange functional \cite{Becke88} with the
gradient-corrected correlation functional of Lee, Yang and Parr
\cite{LYP}. The wave function of all electrons in the system was
expanded using a standard basis set B3LYP/6-31G(2d,p). The PESs calculated within the DFT approach have been
analyzed in Ref.~\cite{ISolovyov06b}.

\begin{figure}[h]
\includegraphics[scale=0.83,clip]{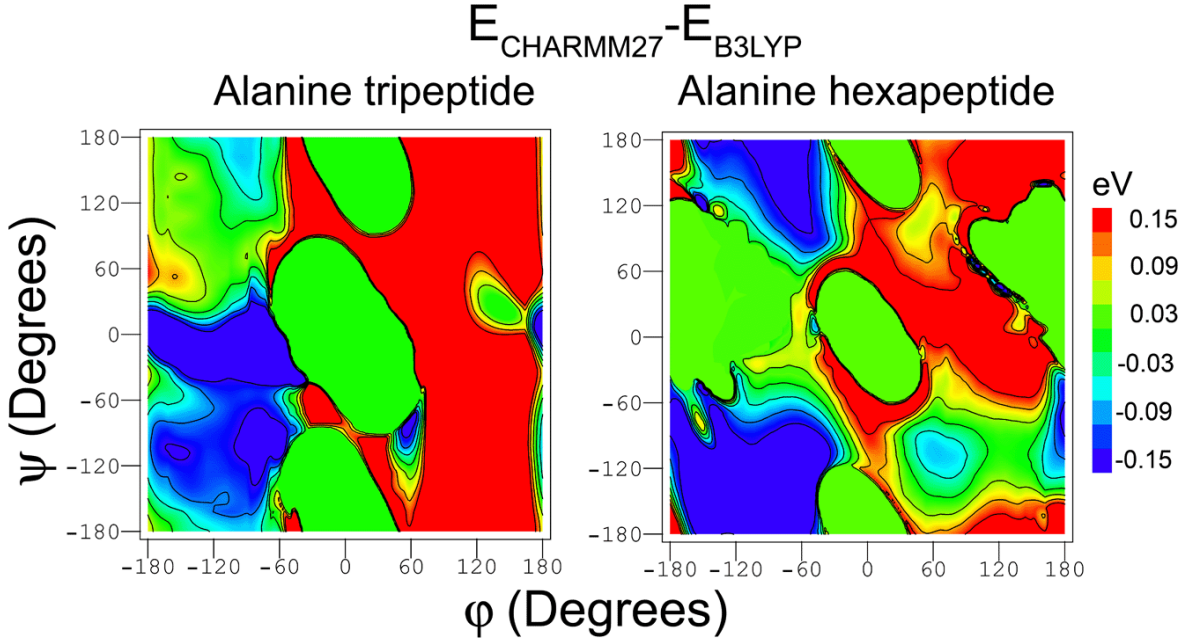}
\caption{Difference between the PESs calculated
with the CHARMM27 force field and with the B3LYP functional
\cite{ISolovyov06b} for the alanine tripeptide (left) and the
alanine hexapeptide (right). The relative energies are given in eV.
The equipotential lines are shown for the energies -0.10, -0.05 0,
0.05 and 0.1 eV.} \label{fg:NAMD_vs_Gaussian}
\end{figure}

The difference between the PESs calculated with
the CHARMM27 force field and with the B3LYP functional is shown in
Fig.~\ref{fg:NAMD_vs_Gaussian} for the alanine tripeptide (left
plot) and for the alanine hexapeptide (right plot).

From Fig.~\ref{fg:NAMD_vs_Gaussian}, we can see that the energy difference
between the PESs calculated with the CHARMM27
force field and with the B3LYP functional is less than 0.15~eV. To
describe the relative deviation of the PESs, we
introduce the relative error of the two methods as follows:

\begin{equation}
\eta=\frac{2\int|E_{B3LYP}(\varphi,\psi)-E_{CHARMM27}(\varphi,\psi)|{\rm
d}\varphi {\rm
d}\psi}{\int|E_{B3LYP}(\varphi,\psi)+E_{CHARMM27}(\varphi,\psi)|{\rm
d}\varphi {\rm d}\psi}\cdot100\%, \label{eq:error}
\end{equation}

\noindent where $E_{B3LYP}(\varphi,\psi)$ and
$E_{CHARMM27}(\varphi,\psi)$ are the potential energies calculated
within the DFT and molecular mechanics methods respectively.
Calculating $\eta$ for alanine tri- and hexapeptide, one obtains:
$\eta_{3\times Ala}=27.6$ \% and $\eta_{6\times Ala}=23.4$ \%
respectively. These values show that the molecular mechanics approach is
reasonable for a qualitative description of the alanine polypeptide. Note however,
that the PES obtained for alanine hexapeptide
within the molecular mechanics method is closer to the PES calculated within the DFT approach. This occurs
because the PESs $E_{CHARMM27}(\varphi,\psi)$
and $E_{B3LYP}(\varphi,\psi)$ of alanine hexapeptide were calculated
for the structure optimized within the DFT approach, while the
PESs $E_{CHARMM27}$ and $E_{B3LYP}$ of alanine
tripeptide were calculated for the structure optimized within the
molecular mechanics method and the DFT approach respectively.

Our analysis shows that the molecular mechanics potential
can be used to describe qualitatively the structural and dynamical
properties of alanine polypeptides with an error of about 20~\%. In
the present paper, we have calculated the thermodynamical properties
of alanine polypeptides with the use of MD method and compared the
obtained results with those attained from the statistical approach.
However, {\it ab initio} MD calculations of alanine polypeptides are
hardly possible on the time scales when the
$\alpha$-helix$\leftrightarrow$random coil phase transition occurs,
even for systems consisting of only 4-5 amino acids~\cite{ISolovyov06b,ISolovyov06c,Yakubovich06b,Yakubovich06c,Salahub01}.
Therefore, we have performed MD simulations for alanine polypeptides
using molecular mechanics forcefield. In order to establish the
accuracy of the statistical mechanics approach, the PES used for the construction of the partition function was also
calculated with the same method.

\subsection{Potential energy surface of alanine polypeptide}
\label{potenergy}

To construct the partition function Eq.~(\ref{Zhc}), one needs to
calculate the PES of a single amino acid in the bounded,
$\epsilon^{(b)}(\varphi,\psi)$, and unbounded,
$\epsilon^{(u)}(\varphi,\psi)$, conformations versus the twisting
degrees of freedom $\varphi$ and $\psi$ (see
Fig.~\ref{fg:angle_def}). The potential energies of alanine in
different conformations determine the $Z_b$ and $Z_u$ contributions
to the partition function, defined in Eqs.~(\ref{Zb})-(\ref{Zu}).

\begin{figure}[h]
\includegraphics[scale=0.85,clip]{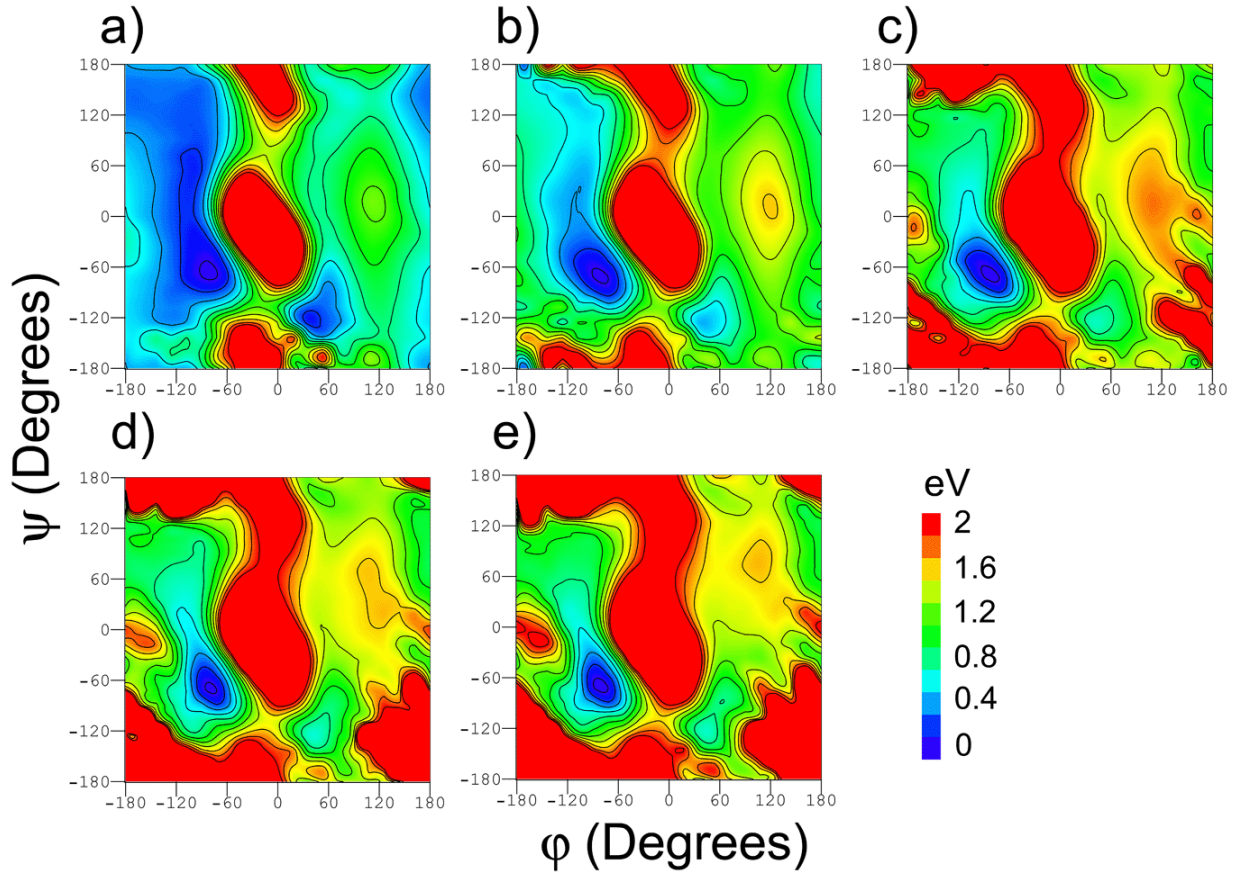}
\caption{PESs for different amino acids of
alanine polypeptide consisting of 21 amino acids calculated as the
function of twisting dihedral angles $\varphi$ and $\psi$ in: a)
second alanine, b) third alanine, c) fourth alanine d) fifth alanine
and e) tenth alanine. Amino acids are numbered starting from the
NH$_2$ terminal of the polypeptide. Energies are given with respect
to the lowest energy minimum of the PES in eV.
The equipotential lines are shown for the energies 1.8, 1.6, 1.4,
1.2, 1.0, 0.8, 0.6, 0.4 and 0.2 eV.} \label{fg:rot2-3-4-5-10}
\end{figure}

The PES of an alanine depends both on the
conformation of the polypeptide and on the amino acid index in the
chain. The PES for different amino acids of the 21-residue
alanine polypeptide calculated as a
function of twisting dihedral angles $\varphi$ and $\psi$ are shown
in Fig.~\ref{fg:rot2-3-4-5-10}. These surfaces were calculated with
the use of the CHARMM27 forcefield for a polypeptide in the
$\alpha$-helix conformation. The PESs a), b),
c), d) and e) in Fig.~\ref{fg:rot2-3-4-5-10} correspond to the
variation of the twisting angles in the second, third, fourth, fifth and
tenth amino acids of the polypeptide respectively. Amino acids are
numbered starting from the NH$_2$ terminal of the polypeptide. We do
not present the PES for the amino acids at boundary
because the angle $\varphi$ is not defined for it.

\begin{figure}[h]
\includegraphics[scale=0.8,clip]{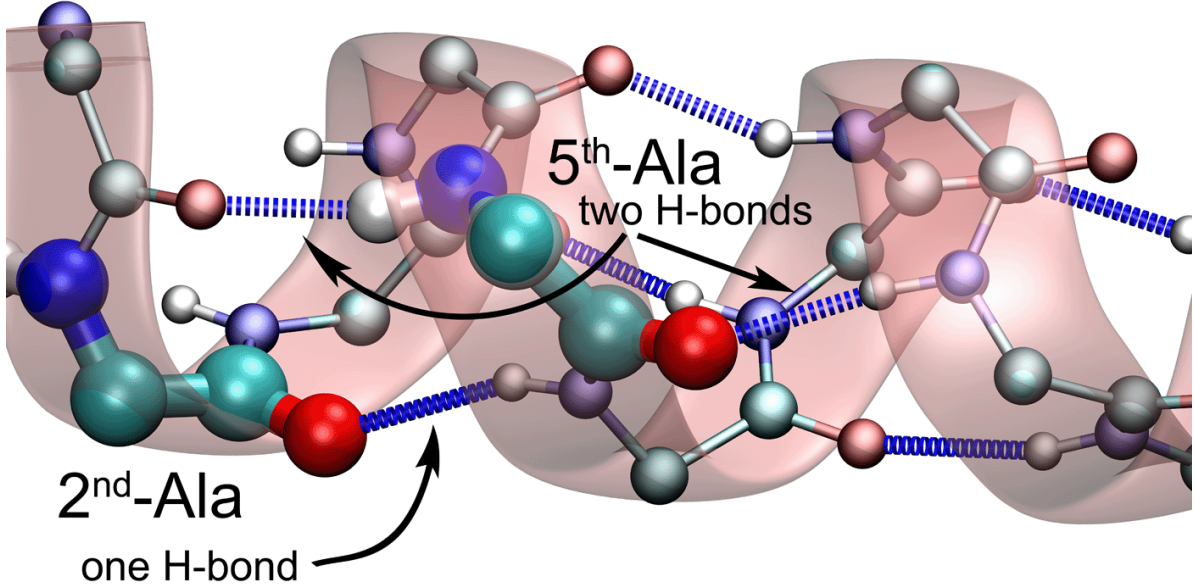}
\caption{Alanine polypeptide in the $\alpha$-helix conformation.
Dashed lines show the hydrogen bonds in the system. Fig. shows
that the second alanine forms only one hydrogen bond, while the
fifth alanine forms two hydrogen bonds with the neighboring amino
acids.} \label{fg:hydrogen_bonds}
\end{figure}

On the PES corresponding to the tenth amino
acid in the polypeptide (see Fig.~\ref{fg:rot2-3-4-5-10}e), one can
identify a prominent minimum at $\varphi=-81^{\circ}$ and
$\psi=-71^{\circ}$. This minimum corresponds to the $\alpha-$helix
conformation of the corresponding amino acid, and energetically,
the most favorable amino acid configuration. In the $\alpha-$helix
conformation the tenth amino acid is stabilized by two hydrogen
bonds (see Fig.~\ref{fg:hydrogen_bonds}). With the change of the
twisting angles $\varphi$ and $\psi$, these hydrogen bonds become
broken and the energy of the system increases. The tenth alanine can
form hydrogen bonds with the neighboring amino acids only in the
$\alpha-$helix conformation, because all other amino acids in the
polypeptide are in this particular conformation. This fact is
clearly seen from the corresponding PES
Fig.~\ref{fg:rot2-3-4-5-10}e, where all local minima have energies
significantly higher than the energy of the global minima (the
energy difference between the global minimum and a local minimum
with the closest energy is $\Delta$E=0.736 eV, which is found at
$\varphi=44^{\circ}$ and $\psi=-124^{\circ}$).

The PES depends on the amino acid index in the
polypeptide. This fact is clearly seen from
Fig.~\ref{fg:rot2-3-4-5-10}. The three boundary amino acids in the
polypeptide form a single hydrogen bond with their neighbors (see
Fig.~\ref{fg:hydrogen_bonds}) and therefore are more weakly bounded than the amino acids inside the
polypeptide. The change in the twisting angles $\varphi$ and $\psi$
in the corresponding amino acids leads to the breaking of these
bonds, hence increasing the energy of the system. However, the boundary
amino acids are more flexible then those inside the polypeptide
chain, and therefore their PES is smoother.
%

Fig.~\ref{fg:rot2-3-4-5-10} shows that the PESs
calculated for the fourth, fifth and the tenth amino acids are very
close and have minor deviations from each other. Therefore, the
PESs for all amino acids in the polypeptide,
except the boundary ones can be considered identical.

Each amino acid inside the polypeptide forms two hydrogen bonds.
However since these bonds are shared by two amino acids, there is
only effectively one hydrogen bond per amino acid (see
Fig.~\ref{fg:hydrogen_bonds}). Therefore, to determine the potential
energy surface of a single amino acid in the bounded,
$\epsilon^{(b)}(\varphi,\psi)$, and unbounded,
$\epsilon^{(u)}(\varphi,\psi)$, conformations, we use the potential
energy surface calculated for the second amino acid of the alanine
polypeptide (see Fig.~\ref{fg:rot2-3-4-5-10}a), because only this
amino acid forms single hydrogen bond with its neighbors (see
Fig.~\ref{fg:hydrogen_bonds}).

The PES of the second amino acid Fig.~\ref{fg:rot2-3-4-5-10}a has a
global minima at $\varphi=-81^{\circ}$ and $\psi=-66^{\circ}$, and
corresponds to the bounded conformation of the alanine. Therefore
the part of the PES in the vicinity of this minima corresponds to
the PES of the bounded state of the polypeptide,
$\epsilon^{(b)}(\varphi,\psi)$. The potential energy of the bounded
state is determined by the energy of the hydrogen bond, which for an
alanine is equal to $E_{HB}=$0.142 eV. This value is obtained from
the difference between the energy of the global minima and the
energy of the plateaus at $\varphi \in (-90^\circ..-100^\circ)$ and
$\psi \in (0^\circ..60^\circ)$ (see Fig.~\ref{fg:rot2-3-4-5-10}a).
Thus, the part of the potential energy surface which has an energy
less then $E_{HB}$ corresponds to the bounded state of alanine,
while the part with energy greater then $E_{HB}$ corresponds to the
unbounded state.

\begin{figure}[h]
\includegraphics[scale=0.8,clip]{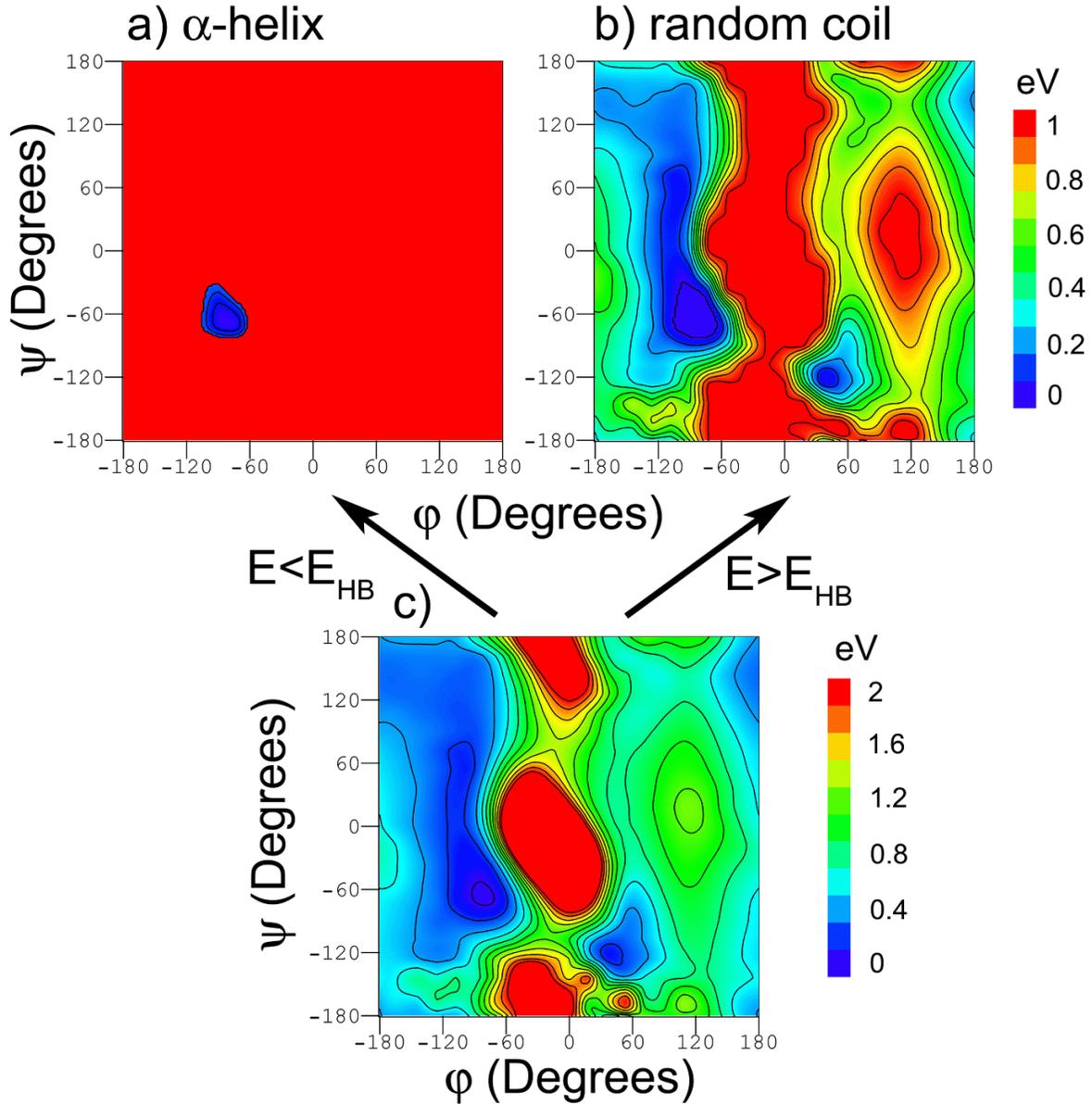}
\caption{PESs for alanine in $\alpha-$helix (plot a) and random coil
conformation (plot b). The potential energy surface for the second
amino acid of the polypeptide is shown in plot c) and is used to
determine the PESs for alanine in $\alpha-$helix and random coil
conformations. The part of the PES shown in plot c, with energy less
then $E_{HB}$ corresponds to the $\alpha-$helix conformation
(bounded state) of the alanine, while the part of the potential
energy surface with energy greater then $E_{HB}$ corresponds to the
random coil conformation (unbounded state). The energies are given
in eV. The equipotential lines in plot a) are shown for the energies
0.05 and 0.1 and 0.15 eV; in plot b) for the energies 0.1, 0.2, 0.3,
0.4, 0.5, 0.6, 0.7, 0.8 and 0.9 eV; in plot c) for the energies 1.8,
1.6, 1.4, 1.2, 1.0, 0.8, 0.6, 0.4 and 0.2 eV.}
\label{fg:pes_helix_coil}
\end{figure}

In Fig.~\ref{fg:pes_helix_coil} we present the potential energy
surfaces for alanine in both the bounded (plot a) and unbounded
(plot b) conformations. Both PESs were calculated from the PES for
the second amino acid in the polypeptide, which is shown in plot c)
of Fig.~\ref{fg:pes_helix_coil}.

\subsection{$\alpha$-helix$\leftrightarrow$random coil phase transition in alanine polypeptide}
\label{phase_trans}

\subsubsection{Internal energy of alanine polypeptide}

Knowing the PES for all amino acids in the polypeptide, one can
construct the partition function of the system using from
Eq.~(\ref{Zhc}). Plots a) and b) in Fig.~\ref{fg:pes_helix_coil}
show the dependence of $\epsilon^{(b)}(\varphi,\psi)$ and
$\epsilon^{(u)}(\varphi,\psi)$ on the twisting angles $\varphi$ and
$\psi$, while $\epsilon^{(b)}$ and $\epsilon^{(u)}$ define the
contributions of the bounded and unbounded states of the polypeptide
to the partition function of the system (see
Eqs.~(\ref{Zb})-(\ref{Zu})). The expressions for $Z_b$ and $Z_u$ are
integrated numerically and the partition function of the polypeptide
is evaluated according to Eq.~(\ref{Zhc}). The partition function
defines all essential thermodynamical characteristics of the system
as discussed in Ref.~\cite{Yakubovich07_preceding}.

The first order phase transition is characterized by an abrupt
change of the internal energy of the system with respect to its
temperature. In the first order phase transition the system either
absorbs or releases a fixed amount of energy while the heat capacity
as a function of temperature has a pronounced peak
\cite{Ptizin_book,Landau05,Prabhu05,Rubin04}. We study the
manifestation of these peculiarities for alanine polypeptide chains
of different lengths.

\begin{figure}[h]
\includegraphics[scale=0.82,clip]{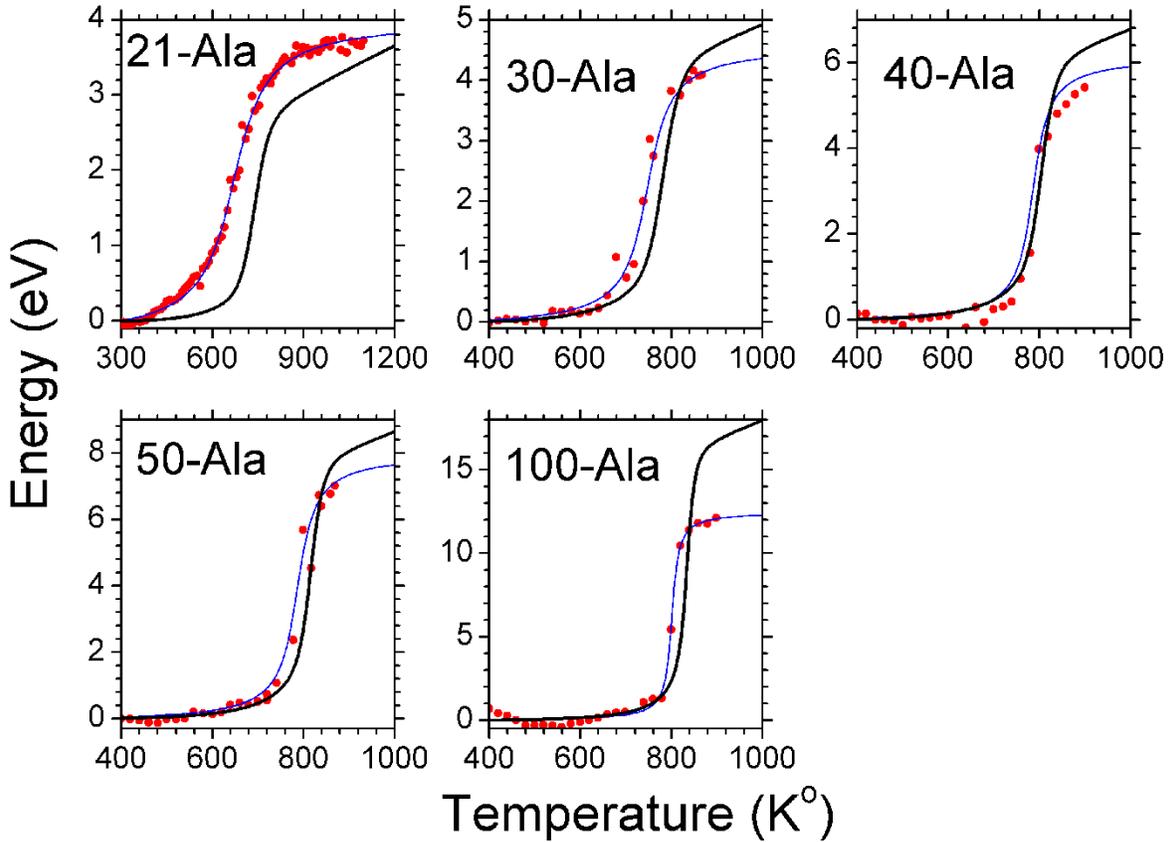}
\caption{Dependencies of the internal energy on temperature
calculated for the alanine polypeptide chains consisting of 21, 30,
40, 50 and 100 amino acids. Thick solid lines correspond to the
results obtained within the framework of the statistical model. Dots
correspond to the results of MD simulations, which are fitted using
Eq.~(\ref{eq:fitting}). The fitting functions are shown with thin
solid lines. The fitting parameters are compiled in
Tab.~\ref{tab:ThermodynamicalParameters}.} \label{fg:trans_energy}
\end{figure}

Fig.~\ref{fg:trans_energy} shows the dependencies of the internal
energy on temperature calculated for alanine polypeptides consisting
of 21, 30, 40, 50 and 100 amino acids. The thick solid lines
correspond to the results obtained using the statistical approach,
while the dots show the results of MD simulations. From
Fig.~\ref{fg:trans_energy} it is seen that the internal energy of
alanine polypeptide rapidly increases in the vicinity of a certain
temperature corresponding to the temperature of the first order
phase transition. The value of the step-like increase of the
internal energy is usually referred as the the latent heat of the
phase transition denoted as $Q$. The latent heat is the
energy that the system absorbs at the phase transition.
Fig.~\ref{fg:trans_energy} shows that the latent heat increases
with the growth of the polypeptide length. This happens because in
the $\alpha$-helix state, long polypeptides have more hydrogen bonds
than short ones and, for the formation of the random coil state,
more energy is required.

The characteristic temperature region of the abrupt change in the
internal energy (half-wight of the heat capacity peak) characterizes
the temperature range of the phase transition. We denote this
quantity as $\Delta T$. With the increase of the polypeptide length
the dependence of the internal energy on temperature becomes steeper
and $\Delta T$ decreases. Therefore, the phase transition in longer
polypeptides is more pronounced. In the following subsection we
discuss in detail the dependence of $\Delta T$ on the polypeptide length.

With the molecular dynamics, one can evaluate the dependence of the
total energy of the system on temperature, which is the sum of the
potential, kinetic and vibrational energies. Then the heat capacity
can be factorized into two terms: one, corresponding to the internal
dynamics of the polypeptide and the other, to the potential energy of the
polypeptide conformation. The conformation of the polypeptide
influences only the term related to the potential energy and
the term corresponding to the internal dynamics is assumed to be
independent of the polypeptides conformation.

This factorization allows one to distinguish from the total energy
the potential energy term corresponding to the structural changes of
the polypeptide. The formalism of this factorization is discussed in
detail in Ref. \cite{Yakubovich07_preceding}. The energy term
corresponding to the internal dynamics of the polypeptide neither
influence the phase transition of the system, nor does it grow
linearly with temperature. The term corresponding to the potential
energy of the polypeptide conformation has a step-like dependence on
temperature that occurs at the temperature of the phase transition.
Since we are interested in the manifestation of the phase transition
we have subtracted the linear term from the total energy of the
system and consider only its non-linear part. The slope of the
linear term was obtained from the dependencies of the total energy
on temperature in the range of 300-450 K$^{\circ}$, which is far
beyond the phase transition temperature (see
Fig.~\ref{fg:trans_energy}). Note that the dependence shown in
Fig.~\ref{fg:trans_energy} corresponds only to the non-linear
potential energy terms.

The heat capacity of the system is defined as the derivative of the
total energy on temperature. However, as seen from
Fig.~\ref{fg:trans_energy} the MD data is scattered in
the vicinity of a certain expectation line. Therefore, the direct
differentiation of the energy obtained within this approach will
lead to non-physical fluctuations of the heat capacity. To overcome
this difficulty we define a fitting function for the total energy of
the polypeptide:

\begin{equation}
E(T)=E_0+\frac{\Delta E}{\pi} \arctan
\left[\frac{T-T_0}{\gamma}\right]+ a T, \label{eq:fitting}
\end{equation}

\noindent where $E_0$, $\Delta E$, $T_0$, $\gamma$ and $a$ are the
fitting parameters. The first and the second terms are related to
the potential energy of the polypeptide conformation, while the last
term describes the linear increase of the total energy with
temperature. The fitting function Eq.~(\ref{eq:fitting}) was used
for the description of the total energy of polypeptides in earlier
papers \cite{Irbaeck04,Irbaeck03}. The results of fitting are shown
in Fig.~\ref{fg:trans_energy} with the thin solid lines. The
corresponding fitting parameters are compiled in
Tab.~\ref{tab:ThermodynamicalParameters}.

\begingroup
\begin{table*}[h]
\caption{Parameters used in Eq.~(\ref{eq:fitting}) to fit the
results of MD simulations.} \label{tab:ThermodynamicalParameters}

\begin{ruledtabular}
\begin{tabular}{cccccc}

     \multicolumn{1}{c}{$n$} &
     \multicolumn{1}{c}{$E_0$} &
     \multicolumn{1}{c}{$\Delta E/\pi$} &
     \multicolumn{1}{c}{$\gamma$} &
     \multicolumn{1}{c}{$T_0$} &
     \multicolumn{1}{c}{$a$}\\

\hline 21 & 11.38$\pm$0.24 & 1.37$\pm$0.10 & 79.4$\pm$7.6 &
670.0$\pm$2.0 & 0.0471$\pm$0.0003\\
30 & 13.61$\pm$0.58 & 1.50$\pm$0.16 & 37.9$\pm$7.3 &
747.4$\pm$3.3 & 0.0699$\pm$0.0008\\
40 & 16.80$\pm$0.39 & 1.991$\pm$0.083 & 26.6$\pm$2.2 &
785.7$\pm$1.8 & 0.0939$\pm$0.0005\\
50 & 19.94$\pm$0.79 & 2.59$\pm$0.21 & 29.4$\pm$5.5 &
786.6$\pm$2.9 & 0.118$\pm$0.0010\\
100& 29.95$\pm$0.67 & 4.00$\pm$0.16 & 10.5$\pm$2.0 &
801.1$\pm$1.1 & 0.2437$\pm$0.0009\\
\hline
\end{tabular}
\end{ruledtabular}
\end{table*}

Fig.~\ref{fg:trans_energy} shows that the results obtained using
the MD approach are in a reasonable agreement with the results
obtained from the the statistical mechanics formalism. The fitting
parameter $\Delta E$ corresponds to the latent heat of the phase
transition, while the temperature width of the phase transition is
related to the parameter $\gamma$. With the increase of the
polypeptides length, the temperature width of the phase transition
decreases  (see $\gamma$ in
Tab.~\ref{tab:ThermodynamicalParameters}), while the latent heat
increases (see $\Delta E$ in
Tab.~\ref{tab:ThermodynamicalParameters}). These features are
correctly reproduced in MD and in our statistical mechanics
approach.

Furthermore, MD simulations demonstrate that with an increase of the polypeptide
length, the temperature of the phase transition shifts towards higher
temperatures (see Fig.~\ref{fg:trans_energy}). The temperature of
the phase transition is described by the fitting parameter $T_0$ in
Tab.~\ref{tab:ThermodynamicalParameters}.  Note also, that the increase
of the phase transition temperature is reproduced correctly within
the framework of the statistical  mechanics approach, as seen from
Fig.~\ref{fg:trans_energy}.

Nonetheless, the results of MD simulations and the results obtained
using the statistical mechanics formalism have several
discrepancies. As seen  from Fig.~\ref{fg:trans_energy} the latent
heat of the phase transition for long polypeptides obtained within
the framework of the statistical approach is higher than that
obtained in MD simulations. This happens because within the
statistical mechanics approach, the potential energy of the
polypeptide is underestimated. Indeed, long polypeptides (consisting
of more than 50 amino acids) tend to form short-living hydrogen
bonds in the random coil conformation. These hydrogen bonds lower
the potential energy of the polypeptide in the random coil
conformation. However, the "dynamic" hydrogen-bonds are neglected in
the present formalism of the partition function construction.

Additionally, the discrepancies between the two methods arise due to
the limited MD simulation time and to the small number of
different temperatures at which the simulations were performed.
Indeed, for alanine polypeptide consisting of 100 amino acids 26
simulations were performed, while only 3-5 simulations correspond to
the phase transition temperature region (see
Fig.~\ref{fg:trans_energy}).

\subsubsection{Heat capacity of alanine polypeptide}

The dependence of the heat capacity on temperature for alanine
polypeptides of different lengths is shown in Fig.~\ref{fg:heat_cap}.
The results obtained using the statistical approach are shown with
the thick solid line, while the results of MD simulations are shown
with the thin solid line. Since the classical heat capacity is constant at
low temperatures, we subtract out this constant value of the
for a better analysis of the phase transition in the
system. We denote the constant contribution to the heat capacity as
$C_{300}$ and calculate it as the heat capacity value at 300
K$^{\circ}$. The $C_{300}$ values for alanine polypeptides of
different length are compiled in the second column of
Tab.~\ref{tab:stat_model}.

\begin{figure}[h]
\includegraphics[scale=0.83,clip]{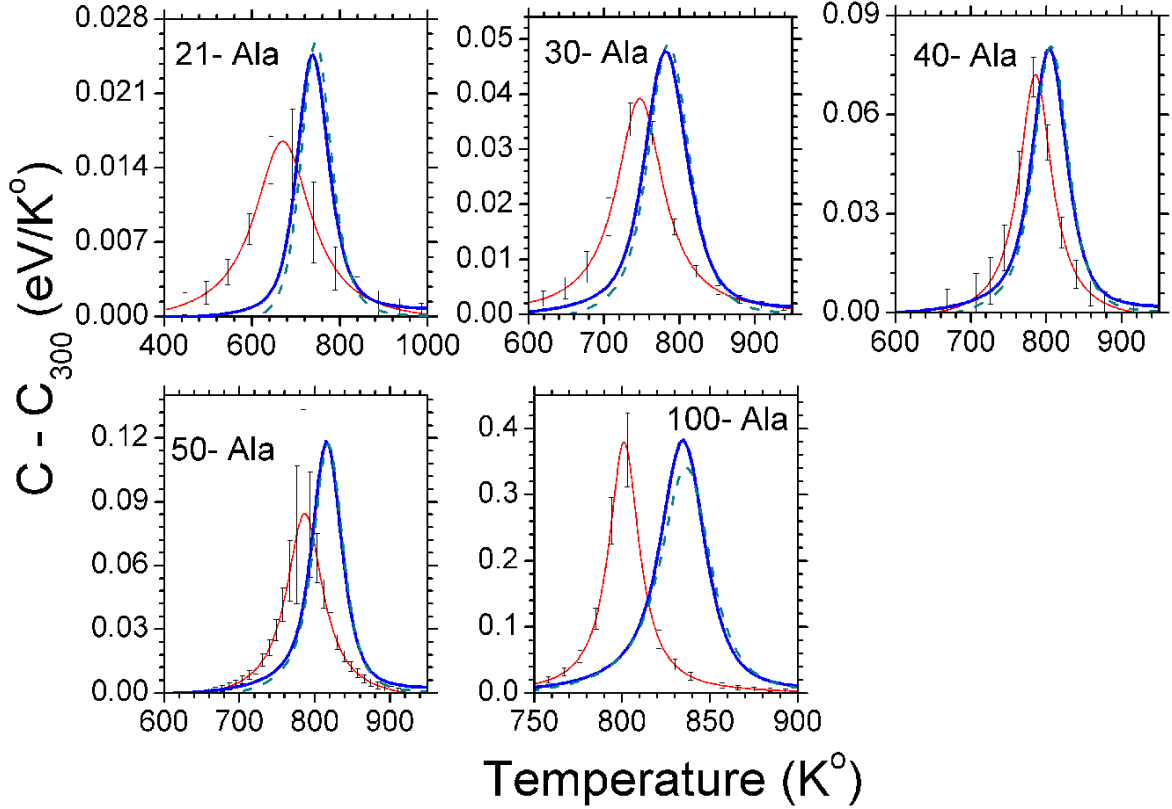}
\caption{Dependencies of the heat capacity on temperature calculated
for the alanine polypeptides consisting of 21, 30, 40, 50 and 100
amino acids. The results obtained using the statistical approach are
shown with the thick solid line, while the results of MD simulations
are shown with the thin solid line. Dashed lines show the heat
capacity as a function of temperature calculated within the
framework of the Zimm-Bragg theory \cite{Zimm59}. $C_{300}$ denotes
the heat capacity at 300 K$^{\circ}$, which are compiled in table
\ref{tab:stat_model}.} \label{fg:heat_cap}
\end{figure}

\begingroup
\begin{table*}[h]
\caption{Parameters, characterizing the heat capacity peak in
Fig.~\ref{fg:heat_cap} calculated using the statistical approach.
Heat capacity at 300 K, $C_{300}$, the transition temperature
$T_{0}$, the maximal value of the heat capacity $C_0$, the
temperature range of the phase transition $\Delta T$ and the
specific heat $Q$ are shown as a function of polypeptide length,
$n$.} \label{tab:stat_model}

\begin{ruledtabular}
\begin{tabular}{ccccccc}

     \multicolumn{1}{c}{$n$} &
     \multicolumn{1}{c}{$C_{300}$ (meV/K)} &
     \multicolumn{1}{c}{$T_{0}$ (K)} &
     \multicolumn{1}{c}{$C_0$ (eV/K)} &
     \multicolumn{1}{c}{$\Delta T$ (K)} &
     \multicolumn{1}{c}{$Q$ (eV)}\\

\hline
21 & 1.951 & 740 & 0.027 & 90 & 1.741\\
30 & 2.725 & 780 & 0.051 & 75 & 2.727\\
40 & 3.584 & 805 & 0.084 & 55 & 3.527\\
50 & 4.443 & 815 & 0.123 & 50 & 4.628\\
100& 8.740 & 835 & 0.392 & 29 & 8.960\\
\hline
\end{tabular}
\end{ruledtabular}
\end{table*}
\endgroup

As seen from Fig.~\ref{fg:heat_cap}, the heat capacity of the system
as a function of temperature acquires a sharp maximum at a certain
temperature corresponding to the temperature of the phase
transition. The peak in the heat capacity is characterized by the
transition temperature $T_{0}$, the maximal value of the heat
capacity $C_0$, the temperature range of the phase transition
$\Delta T$ and the latent heat of the phase transition $Q$. These
parameters have been extensively discussed in our preceding paper
\cite{Yakubovich07_preceding}. Within the
framework of the two-energy level model describing the first order
phase transition, it is shown that:

\begin{eqnarray}
\nonumber T_0&\sim&\frac{{\Delta E}}{\Delta S} = const\\
C_0&\sim&\Delta S^2 \sim n^2\\
\nonumber Q&\sim&\Delta E \sim n\\
\nonumber \Delta T&\sim&\frac{\Delta E}{\Delta S^2} \sim
\frac{1}{n}. \label{eq:parametric}
\end{eqnarray}

\noindent Here $\Delta E$ and $\Delta S$ are the energy and the
entropy changes between the $\alpha-$helix and the random coil
states of the polypeptide, while $n$ is the number of amino acids in
the polypeptide. Fig.~\ref{fg:param_grib} shows the dependence of
the $\alpha$-helix$\leftrightarrow$random coil phase transition
characteristics on the length of the alanine polypeptide. The maximal
heat capacity $C_0$ and the temperature range of the phase
transition $\Delta T$ are plotted against the squared number of amino
acids ($n^2$) and the inverse number of amino acids ($\frac{1}{n}$)
respectively, while the temperature of the phase transition $T_0$
and the latent heat of the phase transition $Q$ are plotted against
the number of amino acids ($n$). Squares and triangles represent the
phase transition parameters calculated using the statistical
approach and those obtained from the MD simulations respectively.
%

\begin{figure}[h]
\includegraphics[scale=0.84,clip]{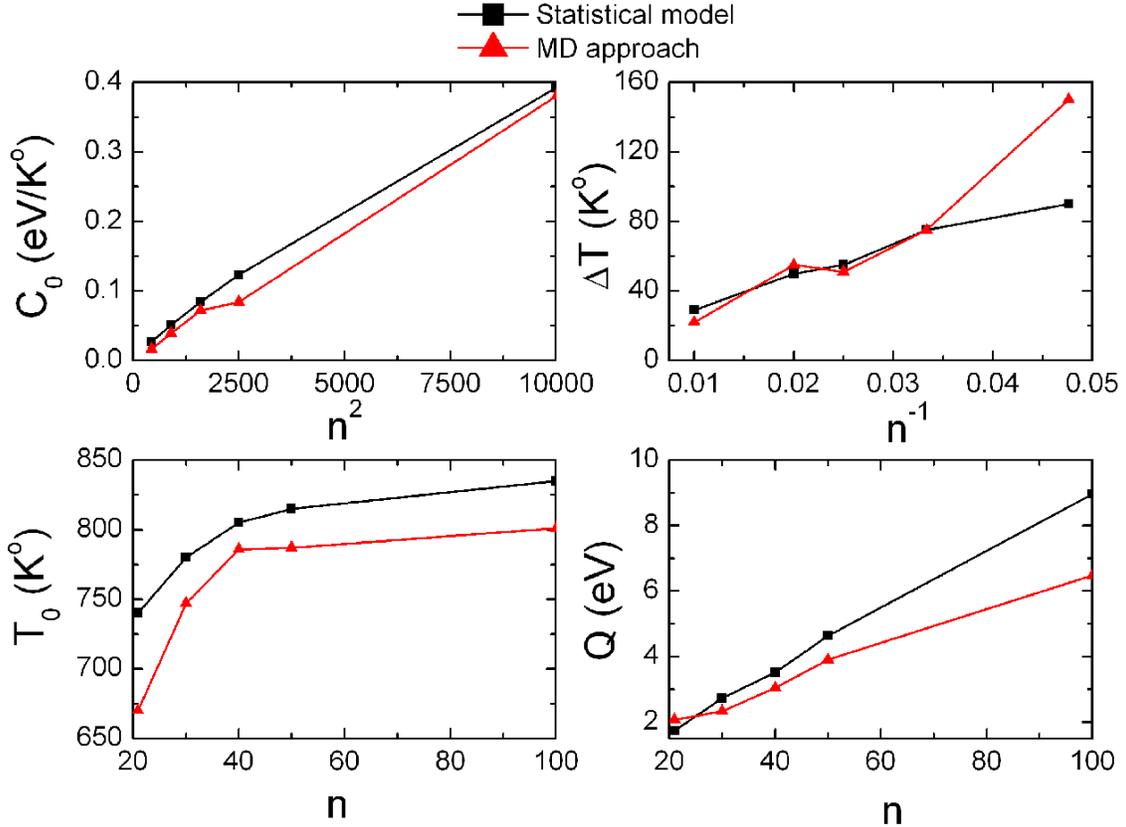}
\caption{Phase transition parameters $C_0$, $\Delta T$, $T_0$ and
$Q$ calculated as a function of polypeptide length. Squares and
triangles represent the phase transition parameters calculated using
the statistical approach and those obtained from the MD simulations
respectively.} \label{fg:param_grib}
\end{figure}

The results obtained within the framework of the statistical model
are in a good agreement with the results obtained on the basis of MD
simulations. The relative deviation of the phase transition
parameters calculated in both methods is on the order of $10\%$ for
short polypeptides and $5\%$ for long polypeptides, as follows from
Fig.~\ref{fg:param_grib}. However, since the MD simulations are
computationally time demanding it is difficult to simulate phase
transition in large polypeptides. The difficulties arise due to the
large fluctuations which appear in the system at the phase
transition temperature and to the large time scale of the phase
transition process. The relative error of the phase transition
temperature obtained on the basis of MD approach is in the order of
$3-5\%$, while the relative error of the heat capacity is about
$30\%$ in the vicinity of the phase transition (see
Fig.~\ref{fg:heat_cap}).

At present, there are no experiments devoted to the study of phase
transition of alanine polypeptides {\it in vacuo}, but such
experiments are feasible and are already planned \footnote{Helmut
Haberland, Private communication.}. In Ref.~\cite{Scheraga70} the
temperature of the $\alpha$-helix$\leftrightarrow$random coil phase
transition was calculated. Depending on the parameter set, the
temperature of the transition ranges from 620 K$^\circ$ to 650
K$^\circ$ for right-handed $\alpha$-helix, and from 730 K$^\circ$ to
800 K$^\circ$ for a left-handed $\alpha$-helix.

In our previous work \cite{Yakubovich06a} on to the
theoretical study of phase transitions in polypeptide chains, we have
introduced the basic ideas of a theoretical method which we
have described in detail in Ref.~\cite{Yakubovich07_preceding} and which we currently apply in this work. It was
shown that the PES calculated as a function of
twisting degrees of freedom $\varphi$ and $\psi$ determines the
partition function of the system. To illustrate our method, we used the PES calculated
for alanine hexapeptide within the framework of the {\it ab initio}
density functional theory\cite{Yakubovich06a} and obtained the
phase transition temperature equal to 300 K$^\circ$. On the other hand, in this paper we established that the phase transition temperature
of alanine polypeptide {\it in vacuo} is 795 K$^\circ$. This is
because in Ref.~\cite{Yakubovich06a} the PES was
calculated for alanine from the hexapeptide. The hydrogen bonds
which stabilize the $\alpha$-helix structure of the hexapeptide are
impaired and therefore the PES of a single
alanine is smoother compared to a long polypeptide where every amino
acid forms two hydrogen bonds. The smoothing of the potential energy
surface results in lowering of the energy barriers and the phase transition temperature.

Nonetheless, smoothing of the PES of an alanine
should happen in solution, as the effective number of hydrogen
bonds in the polypeptide decreases. This fact was demonstrated previously \cite{Yakubovich06a}, where we compared results
of our calculation with available experimental data on alanine rich
peptides in water solution and observed a good correspondence of the
phase transition temperature.

The heat capacity peak is asymmetric. The heat capacity at higher
temperatures, beyond the heat capacity peak, is not zero and forms a
plateau (see Fig.~\ref{fg:heat_cap}). The plateau is formed due to
the conformations of the amino acids with larger energies \cite{Yakubovich06a}. At T=1000 K$^{\circ}$), the difference in
the heat capacity of the polypeptide is $7.6\cdot10^{-4}$,
$1.2\cdot10^{-3}$, $1.6\cdot10^{-3}$, $2.1\cdot10^{-3}$ and
$4.3\cdot10^{-3}$ eV/K$^{\circ}$ for the Ala$_{21}$, Ala$_{30}$,
Ala$_{40}$, Ala$_{50}$ and Ala$_{100}$ peptides respectively. The
magnitude of the plateau increases with the growth of the
polypeptide length. This happens because the number of energy levels
with high energies rapidly increases for longer polypeptide chains.

\subsubsection{Calculation of the Zimm-Bragg parameters}

An alternative theoretical approach for the study of
$\alpha$-helix$\leftrightarrow$random coil phase transition in
polypeptides was introduced by Zimm and Bragg~\cite{Zimm59}. It is
based on the construction of the partition function of a polypeptide
involving two parameters $s$ and $\sigma$, where $s$ describes the
contribution of a bounded amino acid relative to that of an
unbounded one, and $\sigma$ describes the entropy loss caused by the
initiation of the $\alpha$-helix formation.

The Zimm-Bragg theory \cite{Zimm59} is semiempirical because it is parameter dependent. The theoretical method
described in our preceding paper \cite{Yakubovich07_preceding} and
which we use in the present paper is different as it does not
include any parameters and the construction of the partition
function is based solely on the PES of a
polypeptide. Therefore, the construction of our partition function
is free of any parameters, and this is what makes it different from the models
suggested previously. Assuming that the polypeptide has a single
helical region, the partition function derived within the Zimm-Bragg
theory, reads as:

\begin{equation}
Q=1^n+\sigma\sum_{k=1}^{n-3}(n-k-2)s^k, \label{eq:zimm_bragg_simple}
\end{equation}

\noindent where $n+1$ is the number amino acids in the polypeptide,
$s$ and $\sigma$ are the parameters of the Zimm-Bragg theory. The
partition function, which we use in the present paper
Eq.~(\ref{Zhc}) can be rewritten in a similar form:

\begin{equation}
Z=\left[1+\beta
s(T)^3\sum_{k=1}^{(n-1)-3}(n-k-3)s(T)^k\right]\xi(T).
\label{eq:zimm_bragg_our}
\end{equation}

\noindent Here $n$ is the number of amino acids in the polypeptide
and the functions $s(T)$ and $\xi(T)$ are defined as:

\begin{eqnarray}
s(T)&=&\frac{\int_{-\pi}^{\pi}\int_{-\pi}^{\pi}\exp{\left(-\frac{\epsilon^{(b)}(\varphi,\psi)}{k
T}\right)}{\rm d}\varphi {\rm
d}\psi}{\int_{-\pi}^{\pi}\int_{-\pi}^{\pi}\exp{\left(-\frac{\epsilon^{(u)}(\varphi,\psi)}{k
T}\right)}{\rm d}\varphi {\rm d}\psi}\\ \label{eq:zimm_bragg_S}
\xi(T)&=&\left[\int_{-\pi}^{\pi}\int_{-\pi}^{\pi}\exp{\left(-\frac{\epsilon^{(u)}(\varphi,\psi)}{k
T}\right)}{\rm d}\varphi {\rm d}\psi\right]^n
\label{eq:zimm_bragg_chi},
\end{eqnarray}

\noindent where $\epsilon^{(b)}(\varphi,\psi)$ and
$\epsilon^{(u)}(\varphi,\psi)$ are the potential energies of a
single amino acid in the bounded and unbounded conformations
respectively calculated versus its twisting degrees of freedom
$\varphi$ and $\psi$. By comparing Eqs.~(\ref{eq:zimm_bragg_simple})
and (\ref{eq:zimm_bragg_our}), one can evaluate the Zimm-Bragg
parameters as:

\begin{equation}
\sigma(T)=\beta(T)s(T)^3, \label{eq:zimm_bragg_sigma}
\end{equation}

\noindent where $\beta(T)$ is defined in Eq.~(\ref{beta}).

The dependence of the Zimm-Bragg parameters $s$ and $\sigma$ on
temperature is shown in Fig.~\ref{fg:ZB_parameters}a and
Fig.~\ref{fg:ZB_parameters}b respectively. The function $-RT\ln(s)$
grows linearly with an increase in temperature, as seen in
Fig.~\ref{fg:ZB_parameters}a. The zero of this function corresponds
to the temperature of the phase transition in an infinitely long
polypeptide. In our calculation it is 860 K$^\circ$ (see black line
in Fig.~\ref{fg:ZB_parameters}a). Parameter $\sigma$ is shown in the
logarithmic scale and has a maximum at $T=560$ K$^\circ$. Note, that
this maximum does not correspond to the temperature of the phase
transition.

The parameters of the Zimm-Bragg theory were considered in earlier
papers~\cite{Scheraga70,Shental-Bechor05,Nowak07}.
In Fig.~\ref{fg:ZB_parameters}a we present the dependence of
parameter $s$ on temperature calculated in \cite{Scheraga70} (see
squares, triangles and stars in Fig.~\ref{fg:ZB_parameters}b) using
a matrix approach described in Ref.~\cite{Lifson61}. The
energies of different polypeptide conformations were calculated
using the force field described in Ref.~\cite{Ooi67}. Squares, triangles
and stars correspond to three different force field parameter sets
used in Ref.~\cite{Scheraga70}, which are denoted as sets A, B and
C. Fig.~\ref{fg:ZB_parameters}a shows that the results of our
calculations are closer to the results obtained using the parameter
set C. This figure also illustrates that the Zimm-Bragg parameter
$s$ depends on the parameter set used. Therefore, the discrepancies
between our calculation and the calculation performed in
Ref.~\cite{Scheraga70} arise due to the utilization of different force fields.

The Zimm-Bragg parameter $\sigma$ was also calculated in
Ref.~\cite{Scheraga70}. However, it was not systematically studied
for the broad range of temperatures, and therefore we do not plot it
in Fig.~\ref{fg:ZB_parameters}b. In Ref.~\cite{Scheraga70} the parameter
$\sigma$ was calculated only for the temperature of the
$\alpha$-helix$\leftrightarrow$random coil phase transition ranging from 620 K$^\circ$ to 800 K$^\circ$.
In Ref.~\cite{Scheraga70}, it was also demonstrated that parameter $\sigma$ is
very sensitive to the force field parameters, being in the range
$10^{-9.0}-10^{-3.6}$. In our calculation $\sigma=10^{-3.4}$ at 860
K$^\circ$. The dependence of the parameter $\sigma$ on the force
field parameters was extensively discussed in Ref.~\cite{Scheraga70}, where
it was demonstrated that this parameter does not have a strong
influence on the thermodynamical characteristics of phase
transition.

\begin{figure}[h]
\includegraphics[scale=0.78,clip]{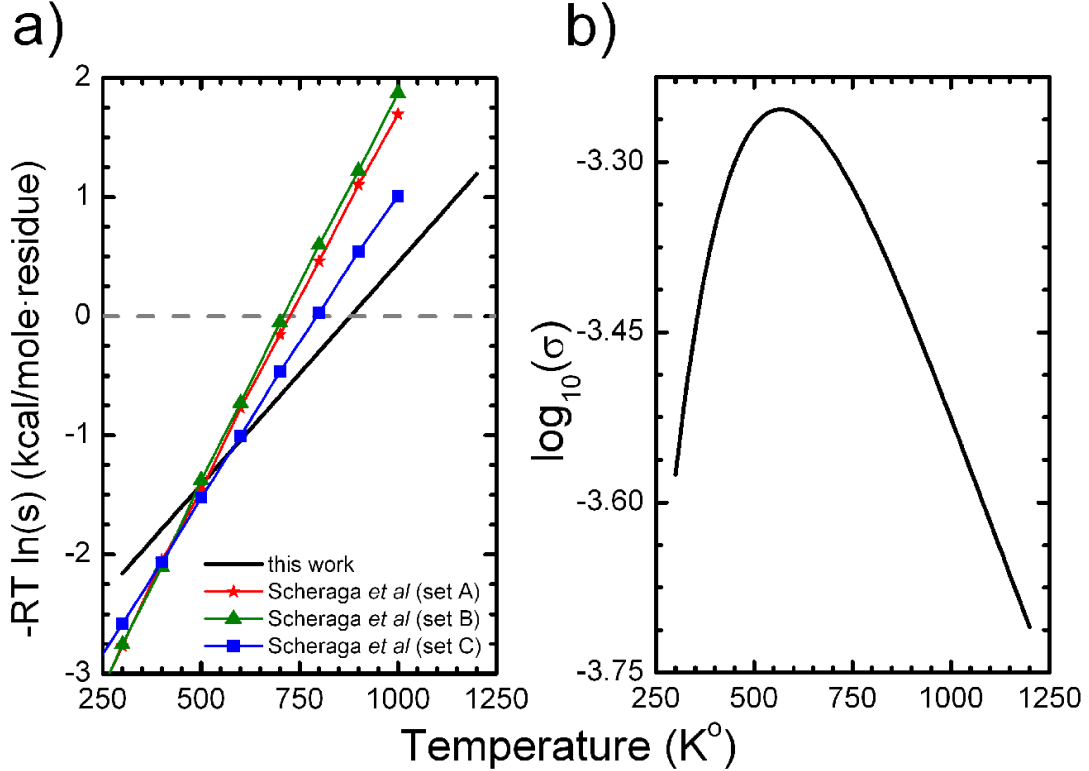}
\caption{Dependence of the parameters of the Zimm-Bragg theory
\cite{Zimm59} $s$ (plot a) and $\sigma$ (plot b) on temperature.
Parameter $s$ describes the contribution to the partition function
of a bounded amino acid relative to that of an unbounded one. The
parameter $\sigma$ describes the entropy loss caused by the
initiation of the $\alpha$-helix formation. Parameter $s$ was also
calculated in Ref.~\cite{Scheraga70} using three different force
fields, shown with stars, triangles and squares in plot a.}
\label{fg:ZB_parameters}
\end{figure}

If the parameters $s$ and $\sigma$ are known, it is possible to construct
the partition function of the polypeptide in the form suggested by
Zimm and Bragg \cite{Zimm59}, and on its basis calculate all
essential thermodynamic characteristics of the system. The
dependence of the heat capacity calculated within the framework of
the Zimm-Bragg theory is shown in Fig.~\ref{fg:heat_cap} by dashed
lines for polypeptides of different length.

From Fig.~\ref{fg:heat_cap} it is seen that results obtained on the
basis of the Zimm-Bragg theory are in a perfect agreement with the
results of our statistical approach. The values of the phase
transition temperature and of the maximal heat capacity in both
cases are close. The comparison shows that the heat capacity
obtained within the framework of the Zimm-Bragg model at
temperatures beyond the phase transition window is slightly lower
than the heat capacity calculated within the framework of our
statistical model.

An important difference of the Zimm-Bragg theory from our theory
arises due to the accounting for the states of the polypeptide with
more than one $\alpha-$helix fragment.
These states are often referred to as multihelical states of the
polypeptide. However, their statistical weight in the partition
function is suppressed. The suppression arises because of entropy loss in the boundary amino acids of a helical fragment. The
boundary amino acids have weaker hydrogen bonds than amino acids in
the central part of the $\alpha$-helix. At the same time the entropy
of such amino acids is smaller than the entropy of an amino acids in
the coil state. These two factors lead to the decrease of the
statistical weight of the multihelical states.

The contribution of the multihelical states to the partition
function leads to the broadening of the heat capacity peak while the
maximal heat capacity decreases. The multihelical states become
important in longer polypeptide chains that consist of more than
100 amino acids. As seen from Fig.~\ref{fg:heat_cap}, the
maximal heat capacity obtained within the framework of the
Zimm-Bragg model for Ala$_{100}$ polypeptide is $10\%$ lower than
that obtained using our suggested statistical
approach. For alanine polypeptide consisting of less than $50$ amino
acids the multihelical states of the polypeptide can be neglected as
seen from the comparison performed in Fig.~\ref{fg:heat_cap}.
Omission of the multihelical states significantly simplifies the
construction and evaluation of the partition function.

\subsubsection{Helicity of alanine polypeptides}

Helicity is an important characteristic of the polypeptide which can
be measured experimentally
\cite{Scholtz91,Lednev01,Thompson97,Williams96}. It describes the
fraction of amino acids in the polypeptide that are in the
$\alpha$-helix conformation. With the increase of temperature the
fraction of amino acids being in the $\alpha-$helix conformation
decreases due to the $\alpha$-helix$\leftrightarrow$random coil
phase transition. In our approach,
the helicity of a polypeptide is defined as follows:

\begin{equation}
\nonumber f_{\alpha}=\frac{\sum_{i=0}^{n-4}(i+1)(n-i-1)Z_u^{i+1}
Z_b^{n-i-1}}{n\left(Z_u^n+\beta \sum_{i=1}^{n-4}(i+1)Z_u^{n+1}
Z_b^{n-i-1}+\beta Z_b^{n-1}Z_u\right)}, \label{eq:helicity}
\end{equation}

\noindent where $n$ is the number of amino acids in the polypeptide,
$Z_b$, $Z_u$ are the contributions to the partition function from
amino acids in the bounded and unbounded states defined in
Eqs.~(\ref{Zb}) and (\ref{Zu}) respectively. The dependence of
helicity on temperature obtained using the statistical approach for
alanine polypeptides of different length are shown in
Fig.~\ref{fg:helicity}.

\begin{figure}[h]
\includegraphics[scale=0.82,clip]{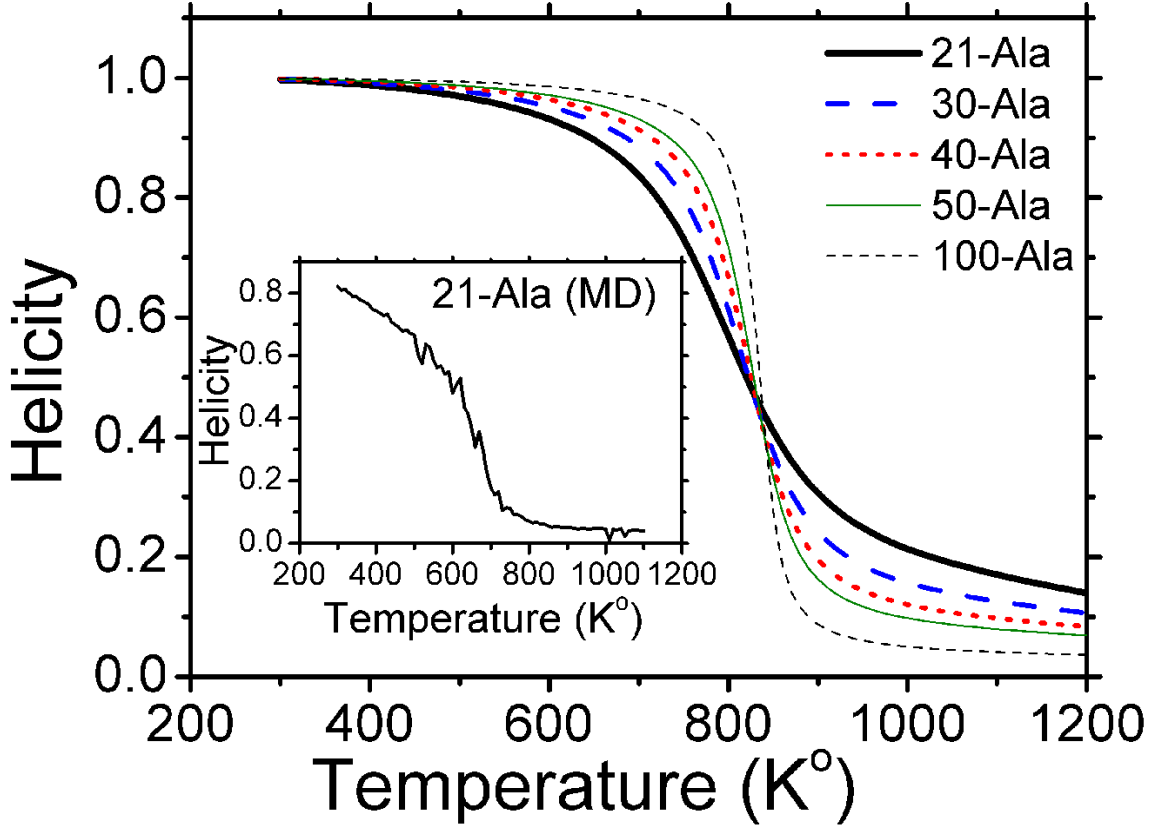}
\caption{Dependency of the helicity on temperature obtained using
the statistical approach for alanine polypeptide chains consisting
of 21, 30, 40, 50 and 100 amino acids. The helicity for alanine
polypeptide consisting of $21$ amino acids obtained within a
framework of MD approach is shown in the inset.} \label{fg:helicity}
\end{figure}

On the basis of MD simulations, it possible to evaluate the
dependence of helicity on temperature. Helicity can be defined
as the ratio of amino acids being in the $\alpha$-helix conformation
to the total number of amino acids in the polypeptide, averaged over
the MD trajectory. The amino acid is considered to be in the
conformation of an $\alpha$-helix if the angles describing its
twisting are within the range of $\varphi\in [-72^{\circ};-6^{\circ}]$ and
$\psi\in[0^{\circ};-82^{\circ}]$. This region was chosen from the
analysis of angles $\varphi$ and $\psi$ distribution at 300
K$^{\circ}$. The helicity for alanine polypeptide consisting of $21$
amino acids obtained within the framework of MD approach is shown in
the inset to Fig.~\ref{fg:helicity}. From this plot it is seen that
at $T\approx300$ K$^{\circ}$, which is far beyond the temperature of
the phase transition, the helicity of the Ala$_{21}$ polypeptide is
0.82. The fact that at low temperatures the helicity of the
polypeptide obtained within the MD approach is smaller than unity
arises due to the difficulty of defining the $\alpha$-helix
state of an amino acid. Thus, the helicity obtained within the MD
approach rolls off at lower temperatures compared to the helicity
of the polypeptide of the same length obtained using the statistical
mechanics approach.

The kink in the helicity curve corresponds to the temperature of the
phase transition of the system. As seen from Fig.~\ref{fg:helicity}, with an increase of the polypeptide length, the helicity curve is
becomes steeper as the phase transition is getting sharper. In
the limiting case of an infinitely long polypeptide chain, the
helicity should behave like a step function. This is yet another feature
of a first-order phase transition.

\subsection{Correlation of different amino acids in the polypeptide}
\label{sec:correlation}

An important question concerns the statistical independence of amino
acids in the polypeptide at different temperatures. In the present
section we analyze how a particular conformation of one amino acids
influences the PES of other amino acids in the
polypeptide. In Fig.~\ref{fg:rmsd} we present the deviations of
angles $\varphi$ and $\psi$ from the twisting angles $\varphi_{10}$
and $\psi_{10}$ in the $10-th$ amino acid of alanine polypeptide.
These results were obtained on the basis of MD simulations of the
Ala$_{21}$ polypeptide at 300 K$^{\circ}$ and at 1000 K$^{\circ}$.
The deviation of angles $\varphi$ and $\psi$ is defined as follows:

\begin{eqnarray}
\label{eq:rmsd}
RMSD(\varphi_i)&=&\sum_{j=1}^{j<=M}\sqrt{\frac{1}{M}(\varphi_i-\varphi_{10})^2}\\
\nonumber
RMSD(\psi_i)&=&\sum_{j=1}^{j<=M}\sqrt{\frac{1}{M}(\psi_i-\psi_{10})^2},
\end{eqnarray}

\noindent where $i$ is the amino acid index in the polypeptide and
$M$ is the number of MD simulation steps. Note, that the plots shown
in Fig.~\ref{fg:rmsd} do not depend on the reference amino acid (we
used the middle amino acid in the polypeptide).

\begin{figure}[h]
\includegraphics[scale=0.85,clip]{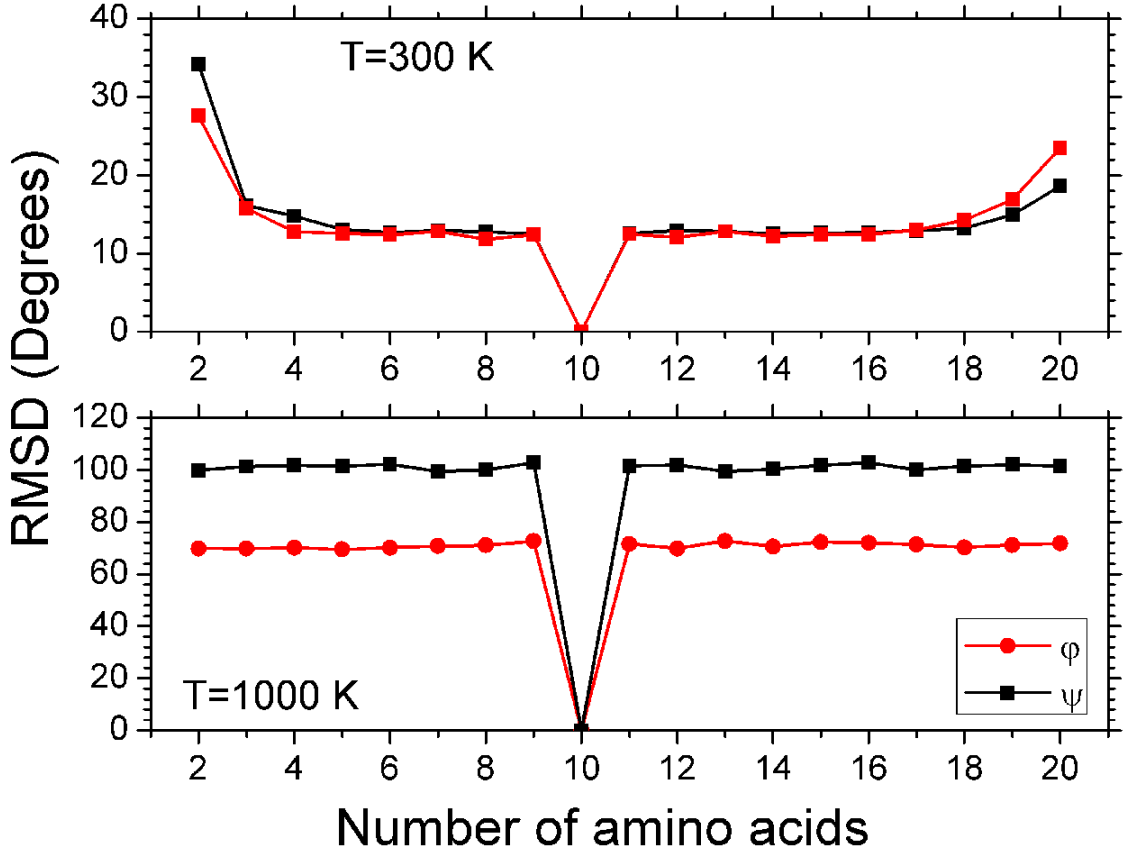}
\caption{The root mean square deviation of angles $\varphi$ and
$\psi$ calculated with the use of Eq.~(\ref{eq:rmsd}) for alanine
polypeptide consisting of 21 amino acids. The calculations were done
in respect to the tenth amino acid of the polypeptide for 300 K (top
plot) and for 1000 K (bottom plot).} \label{fg:rmsd}
\end{figure}

The top plot in Fig.~\ref{fg:rmsd} was obtained at 300 K$^{\circ}$.
At this temperature, all amino acids in the polypeptide are in the
$\alpha-$helix conformation, and the deviation of angles $\varphi$
and $\psi$ is less than 16$^{\circ}$ for all amino acids except the
boundary ones, where the relative deviation of the angles $\varphi$
and $\psi$ is 28$^{\circ}$ and 34$^{\circ}$ respectively. This
happens because, while the boundary amino acids are loosely bounded, the central amino acids in the polypeptide are close
to the minima that corresponds to an $\alpha-$helix conformation.
In the $\alpha-$helix state, all central amino acids are stabilized
by two hydrogen bonds, while the boundary amino acids form only one
hydrogen bond.

At 1000 K$^{\circ}$ the polypeptide is, to large extent, found in the
random coil phase and therefore becomes more flexible. In the
random coil phase, the stabilizing hydrogen bonds are broken, and the
deviation of angles $\varphi$ and $\psi$ significantly increases.
This fact is clearly seen in the bottom plot of Fig.~\ref{fg:rmsd}.
However at 1000 K, the deviation of angles $\varphi$ and $\psi$
in the central and in the boundary amino acids is almost the same,
confirming the assumption that in the random coil phase, short
alanine polypeptides do not build hydrogen bonds.

Another important fact which is worth mentioning is that in the
random coil phase (and in the central part of the $\alpha-$helix),
the deviation of angles $\varphi$ and $\psi$ does not depend on the
distance between amino acids in the polypeptide chain. For instance,
the deviation between angles in the $10-th$ and in the $11-th$ amino
acid is almost the same as the deviation between angles in the
$10-th$ and in the $17-th$ amino acid. This fact allows one to
conclude that in a certain phase of the polypeptide ($\alpha$-helix
or random coil), amino acids can be treated as statistically
independent.

\section{Conclusion}
\label{conclusion}

In the present paper we presented results of calculations obtained
with the statistical method described in our preceding
paper \cite{Yakubovich07_preceding}. We have also performed a detail analysis of the
$\alpha$-helix$\leftrightarrow$random coil transition in alanine
polypeptides of different lengths. We have calculated the potential
energy surfaces of polypeptides with respect to their twisting
degrees of freedom and constructed a parameter-free partition
function of the polypeptide using our statistical formalism
\cite{Yakubovich07_preceding}. From this partition function, we derived and analyzed the temperature dependence of the heat
capacity, latent heat and helicity of alanine polypeptides
consisting of 21, 30, 40, 50 and 100 amino acids. Alternatively, we
have obtained the same thermodynamical characteristics from the use of
molecular dynamics simulations and compared them with the results of our statistical mechanics approach. The comparison proved the
validity of our method and established its
accuracy.

It was demonstrated that the heat capacity of alanine polypeptides has a peak at a certain temperature. The
parameters of this peak (i.e. the maximal value of the heat
capacity, the temperature of the peak, the width at half maximum,
the area of the peak) were analyzed as a function of
polypeptide length. Based on the predictions of the two energy-level
model, it was demonstrated that the
$\alpha$-helix$\leftrightarrow$random coil transition in alanine
polypeptide is a first order phase transition.

We have established a correspondence of our method with the
results of the semiempirical approach suggested by Zimm and Bragg
\cite{Zimm59}. For this purpose we have determined the key
parameters of the Zimm-Bragg semiempirical statistical theory. The calculated
parameters of the Zimm-Bragg theory were compared with the results
of earlier calculations from Ref.~\cite{Scheraga70}.

The final part of this paper deals with the statistical independence
of amino acids in the polypeptide at different temperatures. It was
shown that a particular conformation of one amino acids influences
the PES of other amino acids in the
polypeptide. We demonstrated that in a certain phase, amino acids can be treated as
statistically independent.

In this paper, we demonstrated that the new statistical
approach is applicable for the description of
$\alpha$-helix$\leftrightarrow$random coil phase transition in
alanine polypeptides. However, this method is general and can be used
to study similar processes in other complex molecular systems. For
example, it would be interesting to apply the suggested formalism to
the study of $\beta$-sheet$\leftrightarrow$random coil phase
transition and to the study of non-homogeneous polypeptides (i.e.
consisting of different amino acids). The suggested method can also
be applied to the description of protein folding---an important question left open for further consideration.

In this work we have investigated
$\alpha$-helix$\leftrightarrow$random coil phase transition of
alanine polypeptides {\it in vacuo}. So far there has been done no
experimental work on $\alpha$-helix$\leftrightarrow$random coil
transition in gas phase. Nevertheless, it is important that such experiments
are possible and can be performed using of the techniques MALDI \cite{Karas88,Karas00,Karas03,Wind04} and the ESI mass spectroscopy
\cite{Fenn89,Hvelplund04}. We hope that our theoretical analysis
of the $\alpha$-helix$\leftrightarrow$random coil in alanine
polypeptides {\it in vacuo} will stimulate experimentalists to
verify our predictions.

\section{Acknowledgments}
This work is partially supported by the European Commission within
the Network of Excellence project EXCELL, by INTAS under the grant
03-51-6170. We are grateful to Ms. Adilah Hussein for critical
reading of the manuscript and several suggestions for improvement.
The possibility to perform complex computer simulations at the
Frankfurt Center for Scientific Computing is also gratefully
acknowledged.

\bibliography{journals_short,PhaseTrans_in_Polypeptides}

\end{document}